\documentclass[aps,twocolumn,prb]{revtex4-1}
\usepackage{graphicx,amssymb,amsmath}
\usepackage{graphicx}
\usepackage{dcolumn}
\usepackage{bm}

\newcommand{\bpar}{$B_{||}$}
\newcommand{\bperp}{$B_{\perp}$}
\newcommand{\btot}{$B_{tot}$}

\begin{document}

\title{Spin and the Coulomb Gap in the Half-Filled Lowest Landau Level}

\author{J.P. Eisenstein}
\affiliation{Institute for Quantum Information and Matter, Department of Physics, California Institute of Technology, Pasadena, CA 91125, USA}
\author{T. Khaire}
\altaffiliation[Present address: ]{Argonne National Labs, Argonne, IL 60439, USA}
\affiliation{Institute for Quantum Information and Matter, Department of Physics, California Institute of Technology, Pasadena, CA 91125, USA}
\author{D. Nandi}
\altaffiliation[Present address: ]{Department of Physics, Harvard University, Cambridge, Massachusetts 02138, USA}
\affiliation{Institute for Quantum Information and Matter, Department of Physics, California Institute of Technology, Pasadena, CA 91125, USA}
\author{A.D.K. Finck}
\altaffiliation[Present address: ]{IBM T.J. Watson Research Center, Yorktown Heights, NY 10598, USA}
\affiliation{Institute for Quantum Information and Matter, Department of Physics, California Institute of Technology, Pasadena, CA 91125, USA}
\author{L.N. Pfeiffer}
\affiliation{Department of Electrical Engineering, Princeton University, Princeton, NJ 08544, USA}
\author{K.W. West}
\affiliation{Department of Electrical Engineering, Princeton University, Princeton, NJ 08544, USA}

\date{\today}

\begin{abstract}The Coulomb gap observed in tunneling between parallel two-dimensional electron systems, each at half filling of the lowest Landau level, is found to depend sensitively on the presence of an in-plane magnetic field.  Especially at low electron density, the width of the Coulomb gap at first increases sharply with in-plane field, but then abruptly levels off.  This behavior appears to coincide with the known transition from partial to complete spin polarization of the half-filled lowest Landau level.  The tunneling gap therefore opens a new window onto the spin configuration of two-dimensional electron systems at high magnetic field.

\end{abstract}

\maketitle

\section{Introduction}
In the presence of a large perpendicular magnetic field, Coulomb interactions between electrons confined to a two-dimensional plane compete with disorder in determining the system's physical properties.  In the clean limit interactions dominate and give rise to a wealth of exotic collective states, including both compressible and incompressible quantum liquids, various solid phases, and quantum nematic liquid crystals \cite{perspectives,fradkin10}.  Moreover, the relatively small spin Zeeman energy in typical two-dimensional electron systems (2DES) can be so overwhelmed by these interactions that ground state spin configurations which defy simple Pauli counting rules can be stabilized \cite{halperin83}.  These unusual spin configurations have been detected and studied experimentally by various means, including conventional electrical transport, photo-luminescence and inelastic light scattering, nuclear magnetic resonance, etc. \cite{spin}.

The effects of Coulomb interactions on 2D electron systems at high magnetic field are most dramatically illustrated by the numerous fractional quantized Hall effect states \cite{tsui82} and the emergent gapless metallic phases of composite fermions (CFs) \cite{jain89,halperin93}, all of which exist at specific fractional fillings of the Landau levels created by the magnetic field.  Nevertheless, these exotic phenomena actually represent relatively subtle variations in the strong electronic correlations that exist throughout the high field regime.  For example, experiments \cite{ashoori90,eisenstein92,brown94} have revealed a suppression of the zero bias conductance for electrons tunneling perpendicularly into the 2DES over a wide range of high magnetic fields.  This suppression of the tunneling conductance is observed regardless of whether the 2DES is in a thermodynamically gapped or gapless phase, and extends over a range of voltages about the 2DES Fermi level \cite{eisenstein92,brown94}.  These observations are by now well understood \cite{hatsugai93,he93,johannson93,efros93,varma94,haussmann96,levitov97} to reflect a Coulomb-interaction-induced pseudo-gap in the tunneling density of states.  This pseudo-gap arises from the inability of the correlated 2DES to rapidly relax the charge density defect created by the rapid injection of an electron by tunneling at low energies.  Put another way, at low energies the ($N$+1)-particle states created by tunneling are essentially orthogonal to the ($N$+1)-particle eigenstates of the 2DES.  

In this paper we report experimental observations which indicate that the tunneling Coulomb gap is sensitive to the spin configuration of the 2DES, a possibility not considered in prior theoretical work \cite{hatsugai93,he93,johannson93,efros93,varma94,haussmann96,levitov97}.  Our experiments consist of measurements of the current-voltage characteristics for electrons tunneling between parallel 2D electron systems in semiconductor double quantum wells in the presence of magnetic fields both perpendicular and parallel to the 2D planes.  We focus on the case of the half-filled lowest Landau level (in each 2D layer), both for simplicity and the existence of a well-developed theory \cite{halperin93,he93} for this situation.  We find that the tunneling Coulomb gap increases and then saturates as the parallel field is applied in a manner consistent with the known transition from partial to complete spin polarization of the 2DES.

\section{Experimental}
The samples used in this experiment are GaAs-based semiconductor heterostructures grown by molecular beam epitaxy.  Each contains two 18 nm GaAs quantum wells separated by a 10 nm Al$_{0.9}$Ga$_{0.1}$As barrier layer.  Silicon delta-doping layers, positioned in the thick Al$_{0.32}$Ga$_{0.68}$As layers above and below the double quantum well, populate the lowest subband of each quantum well with a 2DES of nominal density $n\approx5\times10^{10}$ cm$^{-2}$ and low temperature mobility $\mu \approx 10^6$ cm$^2$/Vs.  Standard lithographic methods are used to pattern the 2DES into a mesa structure; for the data presented here this mesa consists of a 250 $\mu$m square with 60 $\mu$m-wide arms extending from each side. Ohmic contacts (NiAuGe) to the individual 2D layers \cite{eisenstein90} are positioned at the ends of these arms. These separate layer contacts enable direct measurements of the interlayer tunneling characteristics of the sample via conventional dc and low frequency ac methods.  The individual layer densities $n_1$ and $n_2$ in the central mesa region are controlled by electrostatic gates deposited on the top and back side of the thinned sample.  We focus here on the balanced case, $n_1=n_2 \equiv n$, with $n$ varied from about $3.9\times 10^{10}$ to $6.3\times 10^{10}$ cm$^{-2}$.   Via {\it in situ} tilting of the sample relative to an applied magnetic field $B_{tot}$, field components both perpendicular (\bperp) and parallel (\bpar) to the 2DES plane could be applied.  Except where otherwise noted, the tunneling data presented here was obtained with \bperp\ adjusted to render the Landau level filling factor $\nu = hn/eB_{\perp} = 1/2$ in each layer \cite{capacitance}.  The density $n$ was kept high enough that condensation into the total filling factor $\nu_T=1/2+1/2=1$ bilayer excitonic phase did not occur. 
 
\begin{figure}
\includegraphics[width=3.1in]{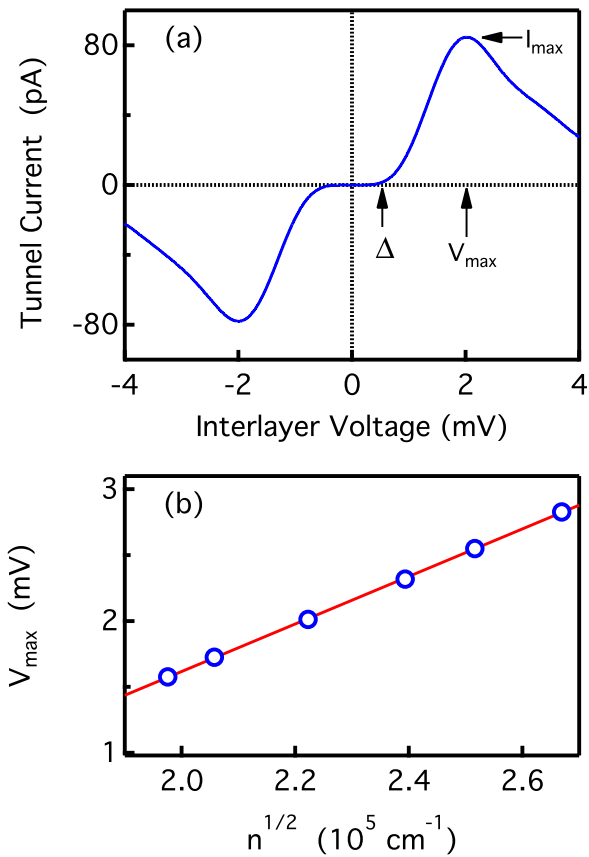}
\caption{\label{}(color online) (a) Typical tunneling $IV$ characteristic at $\nu = 1/2$.  Data taken at $T = 50$ mK with \bperp=4.09 T and \bpar=0. $I_{max}$ and $V_{max}$ denote location of maximum tunneling, while $\Delta$ is voltage where tunneling current has reached 0.02$I_{max}$. (b) Density dependence of $V_{max}$ at \bpar=0.  Solid line is an unweighted least-squares fit of $V_{max}$ versus $n^{1/2}$; the fit extrapolates to $V_{ex}=-2.0$ mV at $n=0$.}
\end{figure}

\section{Results}
Figure 1(a) shows a typical tunneling current-voltage $IV$ characteristic at $\nu = 1/2$.  A pronounced suppression of the tunneling current $I$ around zero interlayer voltage $V$ is readily apparent.  This feature, a Coulomb pseudogap, is the main focus of this paper.  Away from $V=0$ the tunneling current rises and forms a broad peak.  Both the Coulomb gap around $V=0$ and the broad peak at finite $V$ are due to the strong electron-electron interactions which dominate the physics of Landau quantized 2D electron systems.  These features have received substantial theoretical scrutiny \cite{hatsugai93,he93,johannson93,efros93,varma94,haussmann96,levitov97} and are qualitatively well-understood.  We emphasize that these theoretical studies assumed the electron spins were fully polarized by the magnetic field \cite{classical}.

Figure 1(b) presents the density dependence of the voltage $V_{max}$ at which the $\nu = 1/2$ tunneling current reaches its maximum value.  As noted previously \cite{eisenstein95}, if Coulomb interactions within each 2DES dominate the tunneling $IV$ curve, one expects $V_{max}$ (at fixed filling factor) to be proportional to $n^{1/2}$.  As Fig. 1(b) shows, $V_{max}$ is linearly dependent on $n^{1/2}$, but extrapolates to a $negative$ value $V_{ex}$ in the $n=0$ limit.  This negative value, $V_{ex}=-2.0$ mV for the data in Fig. 1(b), represents the excitonic attraction, in the final state, of a tunneled electron and the hole it leaves behind.  In agreement with earlier work, $V_{ex}\approx - 0.5 e^2/\epsilon d$, with $d=28$ nm, the center-to-center spacing between the quantum wells \cite{eisenstein95}.

For a single layer 2DES, adding an in-plane magnetic field \bpar\ to a pre-existing perpendicular field \bperp\ increases the spin Zeeman energy ($E_Z=g\mu_BB_{tot}$, with $g$ the Lande $g$-factor and $\mu_B$ the Bohr magneton) and couples to the finite thickness of the 2DES thereby inducing mixing between Landau levels and the subbands of the confinement potential.  For electrons tunneling between two parallel 2DESs separated by a distance $d$ there is an additional effect arising from the Lorentz force associated with the in-plane field; a tunneling electron acquires a ``momentum boost'' $\hbar q$, with $q= e d B_{||}/\hbar$.  At \bperp=0 this momentum boost can completely suppress the zero bias tunneling conductance if $\hbar q > 2k_F$, with $k_F$ the Fermi wavevector \cite{eisenstein91}.   At high \bperp, with the Fermi level in the lowest Landau level, the momentum boost leads to an exponential suppression of the tunneling matrix element $t$.  Ignoring all other effects of the in-plane field, the tunneling current is expected to follow $I(B_{||})=I(0)~{\rm exp}(-q^2\ell^2/2)$, with $\ell=(\hbar/eB_{\perp})^{1/2}$ the magnetic length \cite{hu92}.  This \bpar-induced suppression of the tunneling current at high \bperp\ is clearly displayed in Fig. 2(a).  The figure plots the maximum tunneling current $I_{max}$ in the broad peak above the Coulomb gap, normalized by its value at \bpar=0, versus $q\ell$. Data for four different 2DES densities, ranging from $n=3.9$ to $n=6.3 \times 10^{10}$ cm$^{-2}$ are shown; in all cases \bperp\ is set to produce $\nu = 1/2$ in each 2DES layer.  Plotted in this way the various data sets collapse onto a single curve and the agreement with theory (dashed line) is excellent.

\begin{figure}
\includegraphics[width=3.1in]{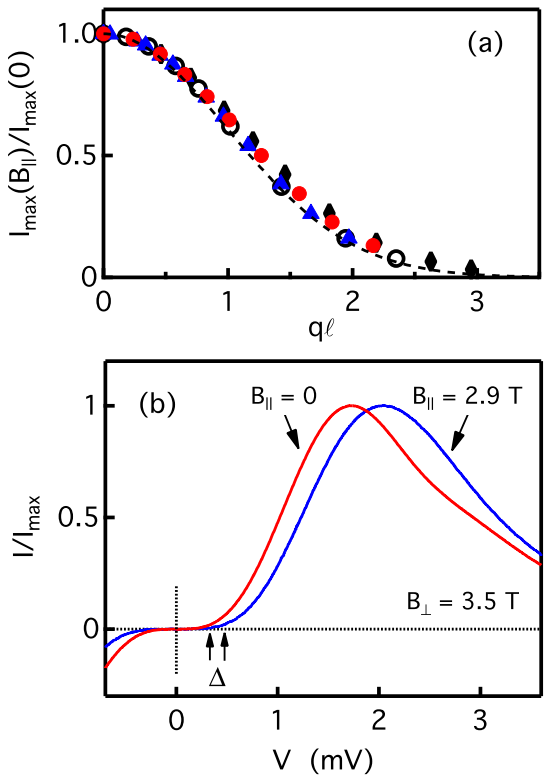}
\caption{\label{}(color online) Effect of an in-plane magnetic field on the tunneling $IV$ curve at $\nu = 1/2$. (a) Maximum tunneling current versus $q \ell$ at various densities.  Dashed line is theory prediction described in text.  (b) Normalized $\nu=1/2$ tunneling $IV$ curves at $n=4.24 \times 10^{10}$ cm$^{-2}$ for \bpar=0 and \bpar=2.9 T.  Upward arrows near lower left show change in $\Delta$ induced by the in-plane field.  Data taken at $T = 50$ mK.}
\end{figure}
If the momentum-boost was the only effect of the in-plane magnetic field then, aside from an amplitude scale factor, the basic $IV$ curve would be independent of \bpar.  We find that this is clearly $not$ the case.  Figure 2(b) presents a typical example of how the normalized tunneling $IV$ curve responds to the application of an in-plane field.  (For these $\nu=1/2$, $n=4.24 \times 10^{10}$ cm$^{-2}$ data, the perpendicular field is fixed at \bperp = 3.51 T.)  Roughly speaking, the entire $IV$ curve expands to higher voltages as \bpar\ is applied.   

Figures 3(a) and 3(b) demonstrate that the \bpar-induced expansion of the $IV$ curve is not a simple rigid shift to higher voltages.  In Fig. 3(a) the voltage location, $V_{max}$, of the peak tunneling current is shown to increase linearly with \btot\ as the in-plane field is applied.  Over the density range studied, we find that the rate of this increase is essentially constant.  

In contrast, the voltage width of the region of strongly suppressed tunneling, $i.e.$ the Coulomb gap, behaves quite differently.   To illustrate this, Fig. 3(b) plots $\Delta$, the voltage at which the tunnel current first reaches 2 percent of its subsequent maximum value, versus \btot.  These data obtain from the same set of tunneling $IV$ curves used to create Fig. 3(a).  In general, $\Delta$ at first rises swiftly as \bpar\ is applied, but then quickly levels off.  For the lowest density data (solid red dots; $n=3.9 \times 10^{10}$ cm$^{-2}$), the initial increase of $\Delta$ is almost two-fold.  As the 2D density is increased, the net increase in $\Delta$ declines until, at $n=6.33 \times 10^{10}$ cm$^{-2}$, little effect remains.  We emphasize that the precise definition of $\Delta$ is not important here, so long as it corresponds to a voltage where the tunneling current is a small fraction of $I_{max}$. 
\begin{figure}
\includegraphics[width=3.1in]{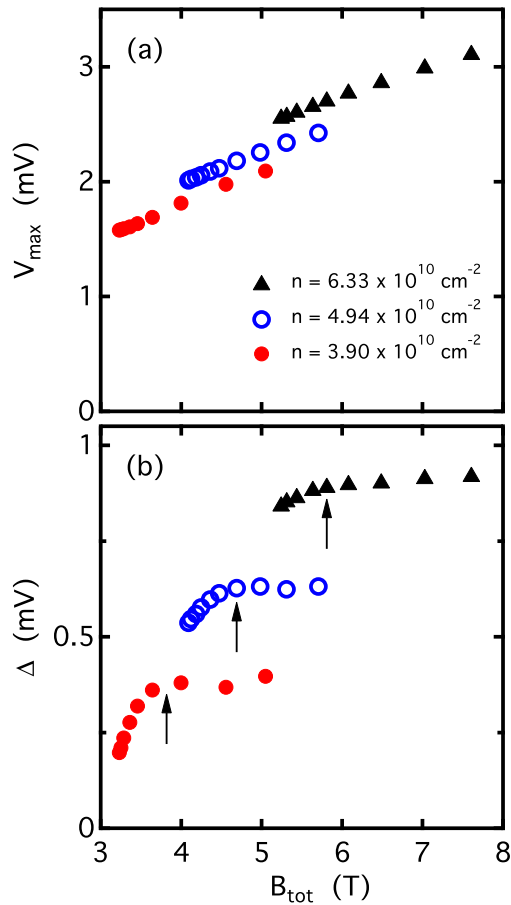}
\caption{\label{}(color online) Effect of in-plane magnetic field on $\nu = 1/2$ tunneling critical points, $V_{max}$ and $\Delta$, at three different densities, plotted versus total magnetic field \btot.  For each density, the perpendicular field \bperp\ is fixed while the in-plane field \bpar\ is varied. The left-most data point of each set corresponds to \bpar=0 (tilt angle $\theta=0$). The arrows indicate our assignment of the ``knee'' in the $\Delta$ $vs.$ \btot\ data.}
\end{figure}

The relatively sharp ``knee'' in the dependence of $\Delta$ on \btot\ suggests that a qualtitative transition in the nature of the 2DES at $\nu =1/2$ occurs as the in-plane field is applied.  Moreover, the transition is most prominent at low density, disappearing almost entirely at the highest densities investigated here.   These observations are at least consistent with a change in the spin configuration of the 2DES, a possibility which we now consider.

\section{Discussion}
There is by now copious evidence that the ground state of the 2DES at $\nu =1/2$ is not fully spin polarized at low density \cite{kukushkin99,dementyev99,melinte00,spielman05,kumada05,dujovne05,tracy07,giudici08,li09,finck10}.  For example, Tracy {\it et al.} \cite{tracy07}, using resistively-detected nuclear magnetic resonance (RDNMR) methods, observed a relatively sudden increase in the nuclear spin lattice relaxation time $T_1$ as the density of a 2DES, maintained at $\nu =1/2$, was increased.   This observation was readily explained by the disappearance of both up and down electronic spin states at the Fermi level as the 2DES transitions from partially to completely spin polarized.  Similarly, Li {\it et al.} \cite{li09}, using both conventional transport and resistively-detected nuclear spin relaxation methods, showed that the $\nu = 1/2$ spin transition could be driven either by increasing the 2DES density or by adding an in-plane magnetic field.  

The composite fermion (CF) theory of the $\nu = 1/2$ state provides a simple way to understand this transition.  In this theory the 2DES at $\nu=1/2$ resembles a Fermi gas at zero magnetic field, only the constituent fermions are CFs, electrons with two fictitious magnetic flux quanta attached.   In general, both up and down spin CFs are present at the Fermi level, their relative populations determined by the comparison between the spin Zeeman energy $E_Z$ and the CF Fermi energy $E_F$.  The Zeeman energy is presumed to be the same as for ordinary electrons, $E_Z = g\mu_B B_{tot}$, with the $g$-factor that appropriate to the host crystal band structure ($|g|=0.44$ for electrons in GaAs).  In contrast, the CF Fermi energy is unrelated to the band structure and, ideally, is determined only by electron-electron interactions.  Dimensional arguments alone then require $E_F=\gamma e^2/\epsilon \ell$, with $\gamma$ a universal, though only approximately known, constant \cite{park98}.  Since $E_Z \propto B_{tot}$ while $E_F \propto B_{\perp}^{1/2}$, it is clear that $E_Z$ can be made to exceed $E_F$ either by increasing the density or by adding an in-plane magnetic field.  Defining the normalized Zeeman energy $\eta=E_Z/(e^2/\epsilon \ell)$, the transition from partial to complete spin polarization at $\nu=1/2$ should occur at $\eta_c=\gamma$, independent of electron density.  Various non-idealities of the 2DES, $e.g.$ the inescapable finite thickness of the 2D layer and interaction-driven Landau level mixing effects, can be expected to disrupt this universality and produce sample-to-sample variations in the observed critical Zeeman energy $\eta_c$ \cite{liu14}.

\begin{figure}
\includegraphics[width=3.1in]{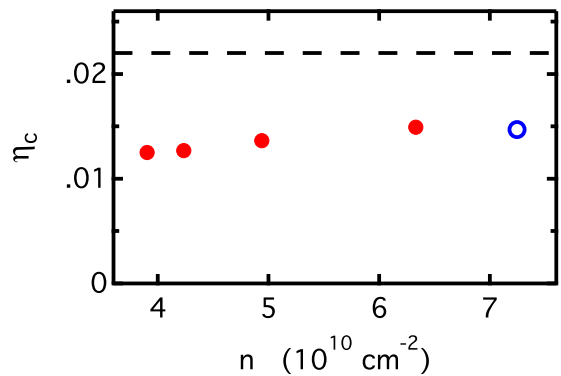}
\caption{\label{}(color online) Solid red circles: Critical normalized Zeeman energy determined from ``knee'' in Coulomb gap $\Delta$ data $vs.$ 2DES density $n$.  Open circle: CF spin transition point found by Tracy {\it et al.} \cite{tracy07}.  Dashed line: Theoretical estimate of Park and Jain \cite{park98}.}
\end{figure}
Using the ``knee'' observed in the $\Delta$ $vs.$ \btot\ data shown in Fig. 3(b) as an indicator of the CF spin transition, the deduced critical Zeeman energy $\eta_c$ is plotted $vs.$ 2DES density $n$ at $\nu=1/2$ in Fig. 4.  Also shown is the $\nu =1/2$ spin transition point found by Tracy {\it et al.} \cite{tracy07} in their RDNMR experiment.  While the agreement between the present experiment and this earlier one strongly supports the identification of the tunneling critical point with the CF spin transition, the near quantitative agreement may be fortuitous.  Both similar \cite{finck10} and somewhat larger \cite{giudici08,li09} values of $\eta_c$ have been observed in other experiments.  The dashed line is the theoretical estimate of Park and Jain \cite{park98}.

To support the identification of the ``knee'' in the $\Delta$ $vs.$ $B_{tot}$ data shown in Fig. 3(b) with the transition from partial to complete spin polarization in the 2DES, the underlying physical mechanism relating these properties of the 2DES needs to be determined.  As is already well-understood \cite{hatsugai93,he93,johannson93,efros93,varma94,haussmann96,levitov97}, the Coulomb gap in the tunneling density of states arises from the electronic correlations created by Coulomb interactions in the Landau quantized 2DES.  These correlations are obviously sensitive to the spin polarization of the 2DES: parallel spin electrons avoid one another more strongly than anti-parallel electrons owing to the Pauli principle.   A partially spin polarized 2DES is in this sense less strongly correlated than a fully polarized one.

A more detailed picture emerges from consideration of the wavevector-dependent conductivity, $\sigma_{xx}(q)$, of the 2DES.  The Coulomb gap itself reflects the inability of the 2DES to rapidly relax the charge defects created by tunneling.  An electron tunneling into (or out of) a Landau quantized 2DES creates a localized excess (or deficit) of charge.  The rate at which these defects can relax to equilibrium is determined by the conductivity, with higher conductivity producing a smaller Coulomb gap and lower conductivity a larger one. This connection between the magnitude of the Coulomb gap and the conductivity was made concrete by He, Platzman, and Halperin (HPH) \cite{he93}.

Hence, we are led to ask how the spin polarization affects the conductivity.  Since at voltages of order $\Delta$ the charge defects created by tunneling are localized on the scale of several magnetic lengths $\ell$, the disorder-dominated $q$=0 conductivity measured in an ordinary electrical transport measurement is not the relevant quantity.  Instead, it is $\sigma_{xx}(q)$ at $q\sim \ell^{-1}$ that counts.  In their study of the CF metal at $\nu =1/2$, Halperin, Lee, and Read \cite{halperin93} calculated the conductivity for the fully spin polarized case: $\sigma_{xx}(q)=(e^2 /8 \pi \hbar)q/ k_F$, where $k_F=(4\pi n)^{1/2}$ is the Fermi wavevector of the spin polarized CF Fermi sea.  This intriguing $q$-linear conductivity was confirmed experimentally in high frequency surface acoustic wave experiments by Willett {\it et al.} \cite{willett90}.

If instead of completely spin polarized, the CF Fermi sea were completely unpolarized, the Fermi wavevector is reduced to $k_F=(2 \pi n)^{1/2}$ and the conductivity $\sigma_{xx}(q)$ thereby {\it increased} \cite{simon_stern} by a factor of $\sqrt{2}$.  Between these extremes, where the 2DES is partially spin polarized, we can expect the conductivity to steadily decrease with increasing spin polarization.  Given the connection between conductivity and the width of the Coulomb gap established by HPH \cite{he93}, this dependence qualitatively explains the behavior of $\Delta$ $vs.$ \btot\ shown in Fig. 3(b) and thus fortifies our association of the ``knee'' in the Coulomb gap data with the transition to complete spin polarization.

\section{Open Questions and Conclusion}
There are, of course, issues that require further study.  For example, as Fig. 3(a) demonstrates, $V_{max}$, the voltage at which the maximum in the tunneling current occurs, does not exhibit any evidence of a transition similar to that displayed by $\Delta$.  We find this behavior unsurprising.  The voltage $V_{max}$ is the cross-over point between two regions of suppressed tunneling.  At low energies the slow dynamics of the correlated 2DES produce the Coulomb gap at the Fermi level.  At high energies, in the single particle cyclotron gap between Landau levels, there are no final states in which to tunnel.  In between these extremes, at energies of order the net Coulomb broadening $E_c \sim e^2/\epsilon \ell$ of the Landau level, the tunneling current has a maximum.  Since $E_c$ depends only on the mean spacing $a$ between electrons ($a=n^{-1/2}=2\pi^{1/2}\ell$ at $\nu = 1/2$), no particular sensitivity to the spin state of these electrons is expected.  The modest \cite{modest} linear increase of $V_{max}$ with $B_{tot}$ shown in Fig. 3(a) nevertheless remains a puzzle. One possibility is that it is due to a stiffening of the effective Coulomb interaction between electrons due to a `squeezing' of their wavefunctions.  Such squeezing can result from the subband/Landau level mixings induced by the in-plane magnetic field.

A more interesting question concerns the role of interlayer Coulomb interactions.  That such interactions exist is made evident by the linear extrapolation to zero density of the $V_{max}$ $vs.$ $n^{1/2}$ data shown in Fig. 1(b).  The substantial negative intercept, $V_{ex} = -2.0$ mV, is believed to be due to a final-state excitonic attraction between a tunneled electron and the hole it leaves behind.  At low energies, {\it e.g.} inside the Coulomb gap, the charge defects created by tunneling are larger in lateral extent \cite{extents} than they are at $V_{max}$.  While this reduces the magnitude of the excitonic attraction, some effect undoubtedly remains.  HPH considered interlayer Coulomb interactions and concluded that in addition to an excitonic `down-shift' of the peak in the tunnel current, the detailed shape of the $IV$ curve deep in the gap was altered \cite{he93}.  To what extent these interactions change the way spin influences the Coulomb gap is not presently known.

A more dramatic interlayer interaction effect is the condensation of the double layer 2DES into an intrinsically bilayer collective state possessing spontaneous interlayer phase coherence \cite{eisenstein14}.  This state, with its remarkable superfluid-like properties, occurs at the same total filling factor as studied here ({\it i.e.} $\nu_T=1/2+1/2=1$) and {\it is} observable in the present samples, but only at densities lower than those explored here.  While some hypothetical precursors of this transition might be influencing the present tunneling results, two considerations suggest otherwise.  First, we have observed that a very similar increase of the Coulomb gap is induced by an in-plane magnetic field when each 2DES is at filling factor $\nu = 0.45$ or $\nu = 0.55$ ($i.e.$ at total filling factors $\nu_T = 0.9$ and 1.1, respectively).  Thus, unlike the coherent bilayer $\nu_T=1$ state, the effects reported here exist over a relatively wide range of filling factors.  Second, we have found that the in-plane field-induced increase in the Coulomb gap persists to at least $T = 0.6$ K. This is well above the temperature at which the bilayer $\nu_T=1$ interlayer coherent state has collapsed \cite{champagne08}.

In conclusion, we find that the Coulomb gap which dominates the low energy tunneling between two parallel 2DESs at high magnetic field can be acutely sensitive to the application of an in-plane magnetic field.  This sensitivity appears to coincide with the transition from partial to complete spin polarization which the in-plane field drives by increasing the spin Zeeman energy.   The wavevector-dependent conductivity of the 2DES provides a plausible link between the Coulomb gap and the system's spin configuration.  The Coulomb gap thus appears to offer a new perspective on the spin state of highly correlated 2D electrons.

\begin{acknowledgements}
It is a pleasure to acknowledge helpful discussions with H.R. Krishnamurthy, P.A. Lee, A.H. MacDonald, and especially S.H. Simon and A. Stern.
This work was supported in part by the Institute for Quantum Information and Matter, an NSF Physics Frontiers Center with support of the Gordon and Betty Moore Foundation through Grant No. GBMF1250.  The work at Princeton University was funded by the Gordon and Betty Moore Foundation through Grant GBMF 4420, and by the National Science Foundation MRSEC Grant 1420541.
\end{acknowledgements}


\begin{references}
\bibitem{capacitance} Strictly true only at zero interlayer voltage $V$.  Capacitive effects at non-zero $V$ disrupt this density balance, but the effects are small and of no relevance here.
\bibitem{classical} Except in Ref. \cite{efros93} where a classical calculation is reported.
\bibitem{modest} The relative increase in $V_{max}$ shown in Fig. 3(a) is about 30\%.  If the excitonic `downshift' (discussed in the context of Fig. 1(b) in the text) is removed from the data, the relative increase is only about 15\%.
\bibitem{extents} Crudely speaking, the size $R$ of the defects is related to the interlayer voltage $V$ through $e^2/\epsilon R \sim eV$.  Hence, smaller $V$ implies larger $R$.
\end{references}
\end{document}